\documentclass[pra,aps,amsmath,amssymb,twocolumn]{revtex4-1}

\usepackage{bm,color,float,dcolumn,amssymb,graphicx,hyperref,subfigure}
\usepackage[titletoc,title]{appendix}

\begin{document}

\title{Emergent Fermi sea in a system of interacting bosons}

\author{Ying-Hai Wu$^{1,2}$ and J. K. Jain$^1$}
\affiliation{$^1$ Department of Physics, The Pennsylvania State University, University Park, PA 16802, USA \\
$^2$ Max-Planck-Institut f{\"u}r Quantenoptik, Hans-Kopfermann-Stra{\ss}e 1, 85748 Garching, Germany}

\date{\today}

\begin{abstract}
An understanding of the possible ways in which interactions can produce fundamentally new emergent many-body states is a central problem of condensed matter physics. We ask if a Fermi sea can arise in a system of bosons subject to contact interaction. Based on exact diagonalization studies and variational wave functions, we predict that such a state is likely to occur when a system of two-component bosons in two dimensions, interacting via a species independent contact interaction, is exposed to a synthetic magnetic field of strength that corresponds to a filling factor of unity. The fermions forming the SU(2) singlet Fermi sea are bound states of bosons and quantized vortices, formed as a result of the repulsive interaction between bosons in the lowest Landau level.
\end{abstract}

\maketitle

\section{Introduction}

While non-interacting fermions form a Fermi sea, an attractive interaction between them can lead to a Bose-Einstein condensation (BEC) of boson-like Cooper pairs or molecules~\cite{Volovik,Regal,Zwierlein,Chen}. One may ask the inverse question: Can interacting bosons produce a Fermi sea? The bound state of multiple bosons cannot behave as a fermion, but examples are known where Fermi-liquid-like physics appears for interacting bosons. A well-known example is that of one-dimensional interacting bosons, whose local properties resemble those of free fermions as manifested in Tonks-Girardeau and Lieb-Liniger systems~\cite{Girardeau,Lieb1,Lenard,Kinoshita,Paredes,Rigol}. The one-dimensional spin-$1/2$ XY model, which is equivalent to a certain model of bosons, can be mapped to a free spinless fermion problem by a Jordan-Wigner transformation~\cite{Lieb2}. For two- and three-dimensional spin systems, free fermionic excitations are thought to emerge in some gapless spin liquids~\cite{Hastings,Ran,Varney}. 

In this article, we explore a different mechanism through which a Fermi sea can emerge in bosonic systems. We consider two-dimensional bosons with repulsive contact interaction and subjected to a synthetic magnetic field. The bosons used in cold atom experiments are charge neutral so do not couple to a real magnetic field, but the effect of a magnetic field can be mimicked either by rapid rotation in a trap~\cite{Ho,Cooper1,Viefers1} or with laser-assisted complex hoppings in optical lattices~\cite{Dalibard,Goldman1}. The single-particle problems of charged particles in continuum and lattice are both well-studied~\cite{Harper1,Wannier,Azbel,Hofstadter}. We shall use the Landau level (LL) eigenstates below but our results are also applicable for a system in which the periodic lattice potential is not very strong~\cite{Harper2}. To characterize the strength of the synthetic magnetic field, we define the ratio between the number of bosons and the number of single-particle states in the lowest Landau level (LLL) as the filling factor $\nu$. In rotating BEC experiments, there has been a significant progress toward bringing a small number of atoms into the strongly correlated regime~\cite{Gemelke}. For lattice systems, a strong uniform magnetic field has been achieved~\cite{Aidelsburger1,Miyake} and the topological invariant (Chern number) has been measured~\cite{Aidelsburger2}.

Previous theoretical works have demonstrated that interacting bosons in synthetic magnetic field can form fractional quantum Hall (FQH) states~\cite{Cooper2,Wilkin,Viefers2,Cooper3,Manninen,Regnault1,Chang,Korslund,Regnault2,Bargi}. The physical mechanism is that the bosons capture one vortex each to turn into composite fermions~\cite{Jain1,Jain15}. The composite fermions experience an effective magnetic field $B^*=B-\rho hc/e$, where $B$ is the actual magnetic field and $\rho$ is the particle density. The integer quantum Hall (IQH) states of composite fermions explains the prominent bosonic FQH states at $\nu=n/(n{\pm}1)$ found in numerical studies~\cite{Regnault1,Chang}. This suggests that the composite fermions might form a Fermi sea when $B^*$ vanishes at $\nu=1$, analogous to the composite fermion (CF) Fermi sea of electrons at half filling of the LLL~\cite{Halperin}. For one-component bosons at $\nu=1$ with contact interaction, it has been shown that even though composite fermions are produced~\cite{Cooper2,Wilkin,Cooper3,Regnault1,Chang}, they experience a weakly attractive residual interaction and form a paired BCS-like state described by the Moore-Read Pfaffian wave function~\cite{Moore}. 

We present below detailed microscopic calculations which strongly suggest that an SU(2) singlet Fermi sea is produced at $\nu=1$ in a system of {\em two-component} bosons interacting with the standard two-body contact interaction. We stress that while our finite system studies below can tell us whether the Fermi sea state is plausible, they cannot prove its existence in a conclusive fashion given the compressible nature of the Fermi sea, and the true test of a Fermi sea will eventually only come from experiments. The experimental techniques available in cold atom systems should enable a preparation of this state and may also be able to measure its many observable consequences. We note that two component bosons in the LLL have been considered in the contexts of FQH effect\cite{Ardonne,Grass1,Furukawa1,Furukawa2,Wu1,Regnault3,Grass2} and of a proposal for the measurement of fractional braiding statistics~\cite{Zhang}. Also, it was proposed by Chung and Jolicoeur \cite{Chung} that a Fermi sea state forms at $\nu=1/3$ for one-component bosons with dipolar interactions. In Appendix A, we show that a Fermi sea may also occur at $\nu=1$ for one-component bosons with long-range two-body and three-body interactions; such interactions are difficult to realize in cold atom experiments.

\section{Models and Methods}

We study bosons on both an open disk and a closed sphere. For disk geometry with the symmetric gauge, the single-particle eigenstates are labeled by a LL index $\alpha{\geq}0$ and an angular momentum index $m{\geq}-\alpha$. The LLL wave functions are
\begin{eqnarray}
\phi_m(z)=\frac{z^m\exp\left(-|z|^2/4\right)}{\sqrt{2\pi 2^m m!}}
\end{eqnarray}
where $z=x+iy$ and $x,y$ are the usual Cartesian coordinates. $\{z^\uparrow\}$ and $\{z^\downarrow\}$ are used to denote respectively the coordinates of spin-up and spin-down particles and $\{z\}$ represent the coordinates of {\em all} particles. In the spherical geometry, the radial magnetic field for particles on a sphere is generated by a monopole at the center of the sphere, whose strength $Q$ has to be an integer or a half integer~\cite{Wu2,Wu3}. The single-particle eigenstates on sphere are labeled by a LL index $\alpha{\geq}0$ and an angular momentum index $\alpha+Q{\geq}m{\geq}-\alpha-Q$. 

The bosons under investigation have two internal state that we refer to as spin-up and spin-down. The number of spin-up particles and the number of spin-down particles are denoted as $N_{\uparrow}$ and $N_{\downarrow}$, respectively. The total number of particles is $N=N_\uparrow+N_\downarrow$. The synthetic magnetic field is assumed to be large enough so the bosons are confined to the LLL and mixing with higher LLs can be neglected. The two-body contact interaction between the particles is encoded in the Hamiltonian
\begin{eqnarray}
H = \sum_{ij} \sum_{\sigma\tau} V_{\sigma\tau} \; \delta^{(2)}(\mathbf{r}^\sigma_i-\mathbf{r}^{\tau}_j) 
\end{eqnarray}
where $i,j$ label the bosons, $\sigma,\tau$ denotes their spins, and $V_{\sigma\tau}$ characterizes the interaction strength (they are determined by the $s$-wave scattering length). In practice, we rewrite this Hamiltonian using the Haldane pseudopotentials~\cite{Haldane} to express it as a sum of projectors which penalize boson pairs with nonzero relative angular momentum. Our main focus below will be on the case in which the two-body contact interaction is SU(2) invariant and we choose $V_{\sigma\tau}$ in such a way that the pseudopotential is one, which will be the units of the energy values presented below. 

The wave function for the emergent CF Fermi sea is
\begin{eqnarray}
\Psi^{\rm CF-FS}_{\rm singlet}(\{z\}) = {\cal P}_{\rm LLL} \biggl[ \Phi^{\rm FS}_{\rm singlet}(\{z^\uparrow\};\{z^\downarrow\}) J(\{z\}) \biggr] 
\label{TwoJainCF}
\end{eqnarray}
Here, $\Phi^{\rm FS}_{\rm singlet}(\{z^\uparrow\};\{z^\downarrow\})$ is the wave function of the spin-singlet Fermi sea in zero magnetic field, and both spin-up and spin-down bosons capture one vortex each to become composite fermions as implemented by multiplying the Jastrow factor $J(\{z\})=\prod^N_{i>j=1} (z_i-z_j)$. In the final step, the wave function is projected to the LLL using the ${\cal P}_{\rm LLL}$ operator.  The wave function $\Phi^{\rm FS}_{\rm singlet}$ has angular momentum $L_z^*=0$, so the $\nu=1$ Fermi sea has angular momentum $L_z=L_z^*+N(N-1)/2=N(N-1)/2$. The CF Fermi sea wave function on sphere, analogous to that in Eq.~\ref{TwoJainCF} for disk, can be obtained using standard methods~\cite{Dev,Wu4,Jain2}. The effective monopole strength for $\Phi^{\rm FS}_{\rm singlet}$ is $Q^*=0$, which implies that the $\nu=1$ Fermi sea occurs at $Q=Q^*+N-1=N-1$. 

\begin{figure}
\includegraphics[width=0.45\textwidth]{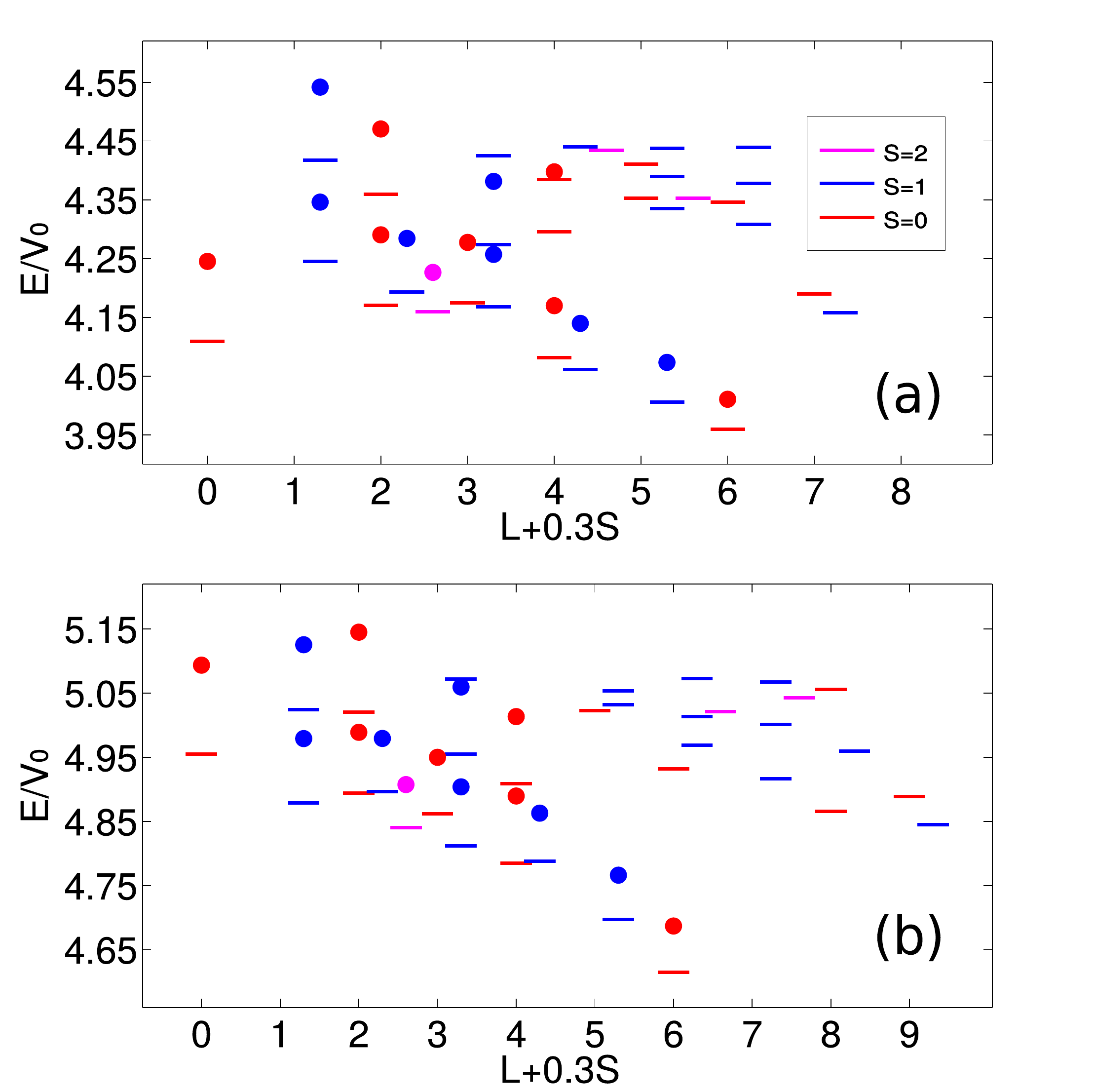}
\caption{ (color online) Energy spectra of two-component bosons with SU(2) invariant contact interaction on sphere (dashes). The dots represent the energies of the CF trial states $\Psi^{\rm CF-FS}_{\rm singlet}$. The values of $(N_{\uparrow},N_{\downarrow},2Q)$ are $(6,6,11)$ in panel (a) and $(7,7,13)$ in panel (b). The horizontal coordinate is the linear combination $L+0.3S$ so the states with the same $L$ but different $S$ can be distinguished by their relative shifts. The spin quantum numbers are also shown in different colors as indicated by the inset of panel (a).}
\label{Figure1}
\end{figure}

\begin{table*}
\centering
\begin{tabular}{|c|ccccccccccc|}
\hline
       &       &        &        &        &        &$(L,S)$ &        &        &        &        &       \\\hline
Figure & (0,0) & (1,1)  & (2,0)  & (2,1)  & (2,2)  & (3,0)  & (3,1)  & (4,0)  & (4,1)  & (5,1)  & (6,0) \\\hline
 1(a)  & 0.931$^\dagger$ & 0.876 & 0.902  & 0.935  & 0.969  & 0.939  & 0.937  & 0.919  & 0.953  & 0.965  & 0.982 \\
       & (1210)  & (7963)   & (5427)   & (12859)  & (14323)  & (7196)   & (17993)  & (9345)   & (22452)  & (27008)  & (12732) \\\hline
 1(b)  & 0.872$^\dagger$ & 0.933 & 0.929  & 0.952  & 0.971  & 0.942  & 0.946  & 0.926  & 0.962  & 0.968  & 0.963 \\
       & (18607) & (136181) & (89876)  & (224106) & (261967) & (123338) & (312212) & (157877) & (395292) & (477017) & (220148)\\\hline   
\end{tabular}
\caption{The overlaps between CF trial states and the corresponding exact eigenstates shown in Fig.~\ref{Figure1}. $L$ is the total angular momentum quantum number and $S$ is the total spin quantum number. When ${\cal N}>1$ multiplets occur in a given $(L,S)$ sector, the overlap is defined as $\sqrt{\frac{1}{\cal N} \sum_{\alpha,\beta} \left[\langle\Psi^{\rm CF-FS}_{{\rm singlet},\alpha}|\Psi^{\rm exact}_\beta\rangle\right]^2}$ where the summation is over all the ${\cal N}$ states. The dagger indicates that there are two CF states in the $(L,S)=(0,0)$ sector but one of them is pushed up to a very high energy [higher than the range of Fig.~\ref{Figure1} (a) and (b)] and we keep only the lower energy state in our comparison. The total number of linearly independent $(L,S)$ multiplets is given in parentheses below each overlap.}
\label{Table1}
\end{table*}

\section{Results and Discussions}

We first show our results for the spherical geometry. The exact energy spectra of two-component bosons on sphere are shown in Fig.~\ref{Figure1}. The energy eigenstates are classified by their total angular momentum $L$ and total spin $S$. The energy levels with different $S$ at the same $L$ are shifted horizontally for clarity. To compare with the CF theory, we take all the possible fermionic wave functions $\Phi^{\rm FS}_{\rm singlet}(\{z^\uparrow\};\{z^\downarrow\})$ at $Q^*=0$, in the basis with definite $L$ and $S$ quantum numbers, for which the composite fermions have lowest effective kinetic energy. When there are multiple linearly independent states in a given $(L,S)$ sector, we diagonalize the Hamiltonian in this reduced basis. The CF predictions thus obtained are also shown in Fig.~\ref{Figure1} for comparison. The overlaps between the CF and the exact states are shown in Table~\ref{Table1}. [For the $(L,S)=(0,0)$ sector, the CF theory nominally predicts two states in both cases, but one of them is pushed to a very high energy so is not shown in Fig.~\ref{Figure1} and Table~\ref{Table1}.] When the system size increases from $N=12$ to $14$, the number of states in each $(L,S)$ sector increases many fold but the overlaps remain large. One can see that the CF Fermi sea description captures the low-energy physics of the systems very accurately. It has been noted that the Hund's rule can be applied with some success to composite fermions~\cite{Jain3,Wu5,Rezayi}. We see from Fig.~\ref{Figure1} that for a given total spin $S$ the energies generally decrease with increasing $L$ as expected based on the Hund's rule. If the Hamiltonian is tuned away from the SU(2) invariant point, the agreement between exact eigenstates and the trial wave functions rapidly worsens, indicating that an SU(2) invariant interaction is optimal for producing the CF Fermi sea.

\begin{figure}
\includegraphics[width=0.4\textwidth]{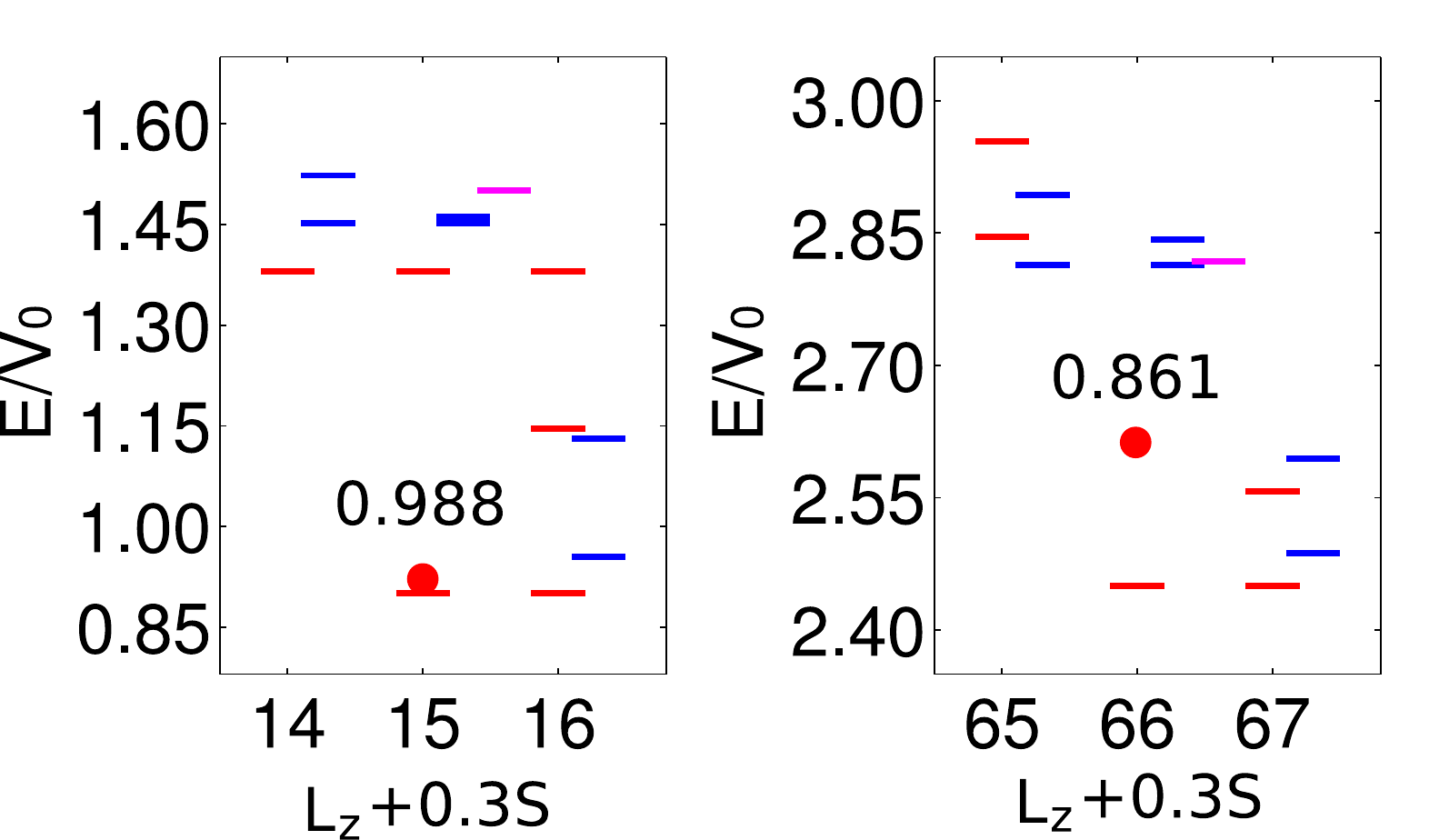}
\caption{ (color online) Energy spectra of two-component bosons with SU(2) invariant contact interaction on disk (dashes). The number of particles is $N=6$ in panel (a) and $N=12$ in panel (b). The horizontal coordinate is the linear combination $L_z+0.3S$ so the states with the same $L_z$ but different $S$ can be distinguished by their relative shifts. The spin quantum numbers are also shown in different colors as in Fig.~\ref{Figure1}. There is a clear cusp at $L_z=N(N-1)/2$, with a spin singlet ground state. The dot shows the energy of the CF wave function, and the number above the dot is the overlap between this wave function and the exact ground state.}
\label{Figure2}
\end{figure}

\begin{figure}
\includegraphics[width=0.45\textwidth]{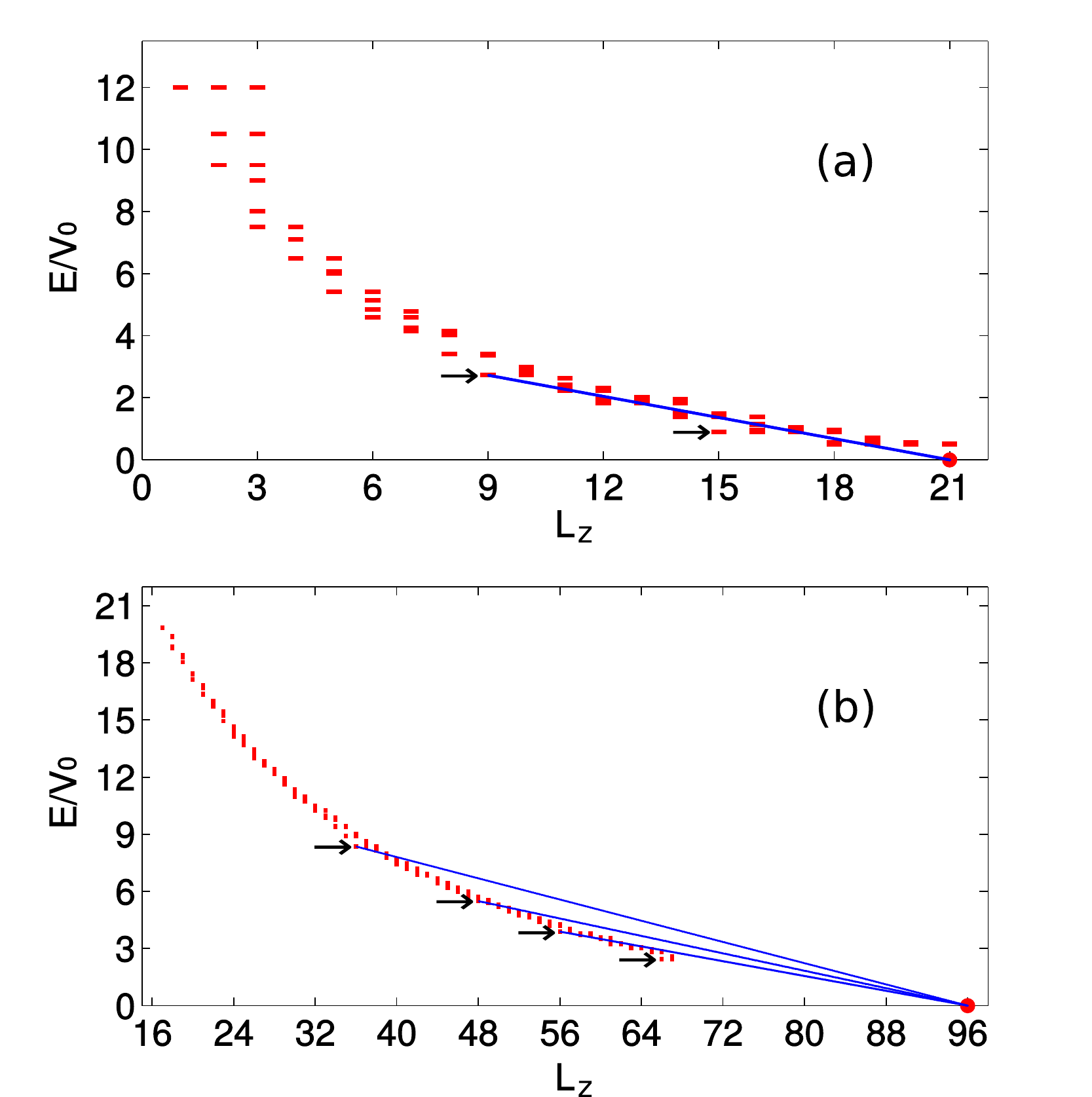}
\caption{ (color online) Energy spectra two-component bosons with SU(2) invariant contact interaction on disk. The number of particles is $N=6$ in panel (a) and $N=12$ in panel (b). For simplicity, the spin quantum numbers are not shown. Some of the  compact states where the energies have a downward cusp are indicated by arrows. The $L_z=15$ state in panel (a) and the $L_z=66$ state in panel (b) are the ones that would evolve into the CF Fermi sea in the thermodynamic limit. The dots mark the Halperin $221$ state which is the maximum density zero energy eigenstate.}
\label{Figure3}
\end{figure}

The cold atom experiments are generally performed in parabolic traps, making it important to understand how the SU(2) singlet Fermi sea would manifest itself in finite open systems on a disk. In this case, the good quantum numbers are the $z$-component of angular momentum $L_z$ and the total spin $S$. This problem is similar to that of interacting electrons in a parabolic quantum dot, which can be well understood using CF theory~\cite{Rejaei1,Rejaei2,Jain4,Jeon1,Jeon2}. The single-particle orbitals of composite fermions are labeled by the CF LL index $\alpha$ and angular momentum index $m$, in analogy to the labels of the single-particle states of the physical bosons. The most relevant many particle states are the ones that appear at downward cusps in the plot of energy versus $L_z$. These correspond to the so-called ``compact states" of composite fermions~\cite{Jain4} denoted as $[N_0,N_1,N_2\cdots]$, wherein every CF LL is compactly occupied (without leaving any holes) with the occupation numbers satisfying $N_0{\geq}N_1{\geq}N_2\cdots$. To have a unique configuration with lowest CF cyclotron energy at $L_z^*=0$ [$L_z=N(N-1)/2$], we fill the innermost $N_\alpha/2$ orbitals in each CF LL according to a pyramid-like structure with $N_{\alpha+1}=N_{\alpha}+2$. As an example, the 12 particle CF Fermi sea is given by the compact state $[6,4,2]$ in which the $(\alpha,m)=(0,0),(0,1),(0,2),(1,-1),(1,0),(2,-2)$ CF orbitals are doubly occupied. We can ask two questions: If we plot the exact energy as a function of $L_z$, is there a downward cusp at $N(N-1)/2$? If so, is it well described by the CF wave function? The results presented in Fig.~\ref{Figure2} show that there indeed are cusps at $L_z=N(N-1)/2$, the ground states at the cusps are spin-singlets, and they have a high overlap with the CF Fermi sea state. To reduce the computational effort, we impose in our numerical study a cutoff on the single-particle angular momentum. For the systems studied here, we find that $m\leq 16$ is a suitable cutoff; we have tested that the many-body eigenvalues remain essentially unchanged when the cutoff is increased to $18$. For the $N=12$, the Hilbert space dimension at $L_z=66$ is 35027058 (with the cutoff), indicating that the overlap $0.861$ is still very significant.

Another important question is whether the CF Fermi sea state can become the global ground state for some range of rotation frequency. If so, this state can be adiabatically prepared as in Ref.~\onlinecite{Gemelke}. To this end, we need to calculate the ground state energy as a function of $L_z$, i.e. the yrast spectrum. For a parabolic confinement potential (with a strength characterized by $\omega_c$), the total energy at rotation frequency $\Omega$ has an additional term $(\omega_c-\Omega)L_z$. If the finite size representation of the CF Fermi sea at $L_z=N(N-1)/2$ is below the line joining any two yrast states, it will become the global ground state for a suitable choice of confinement potential. For the SU(2) invariant two-body contact interaction, the Halperin 221 wave function is the maximum density zero energy solution, which occurs at total angular momentum $L^{\rm 221}_z=N(3N-4)/4$~\cite{Halperin2}, so it is sufficient to obtain the yrast spectrum up to $L^{\rm 221}_z$. We show in Fig.~\ref{Figure3} the energy spectra of the $N=6$ and $12$ systems at many different angular momenta. For the $N=6$ system shown in Fig.~\ref{Figure3} (a), we are able to calculate the low-lying eigenvalues from $L_z=0$ to $L_z=21$ and the CF Fermi sea indeed can become the global ground state for some $\Omega$. For the $N=12$ system shown in Fig.~\ref{Figure3} (b), we are only able to obtain the spectrum up to $L_z=67$ and the CF Fermi sea can be tuned to the global ground state compared to the states at $L_z<67$. One expects to see cusp states at $L_z=72,76,84$ but we are unable to calcualte their energies. 

\begin{figure}
\includegraphics[width=0.5\textwidth]{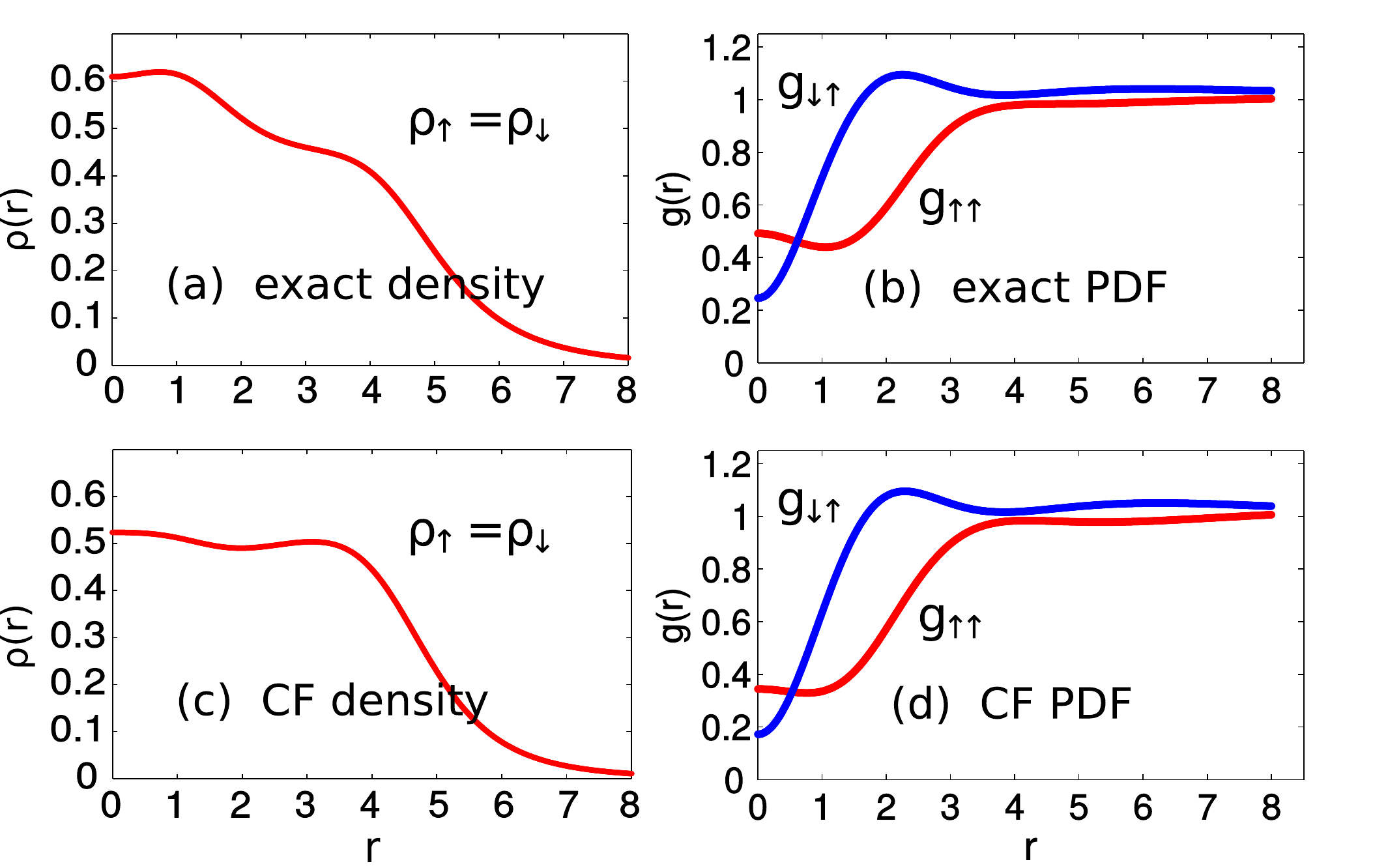}
\caption{(color online) Density profiles and pair distribution functions of the exact ground state and the CF trial state for the $N=12$ system on disk. The distance $r$ is measured in units of the magnetic length $\ell_B=\sqrt{{\hbar}c/eB}$. As we restrict $m\leq16$ in exact diagonalizations, we have normalized the density profiles by that of a uniform state with total filling $\nu=1$. In panels (b) and (d), the two lines show $g_{\uparrow\uparrow}$ and $g_{\downarrow\uparrow}$ as indicated by the symbols in their vicinities.}
\label{Figure4}
\end{figure}

\begin{figure*}
\includegraphics[width=0.9\textwidth]{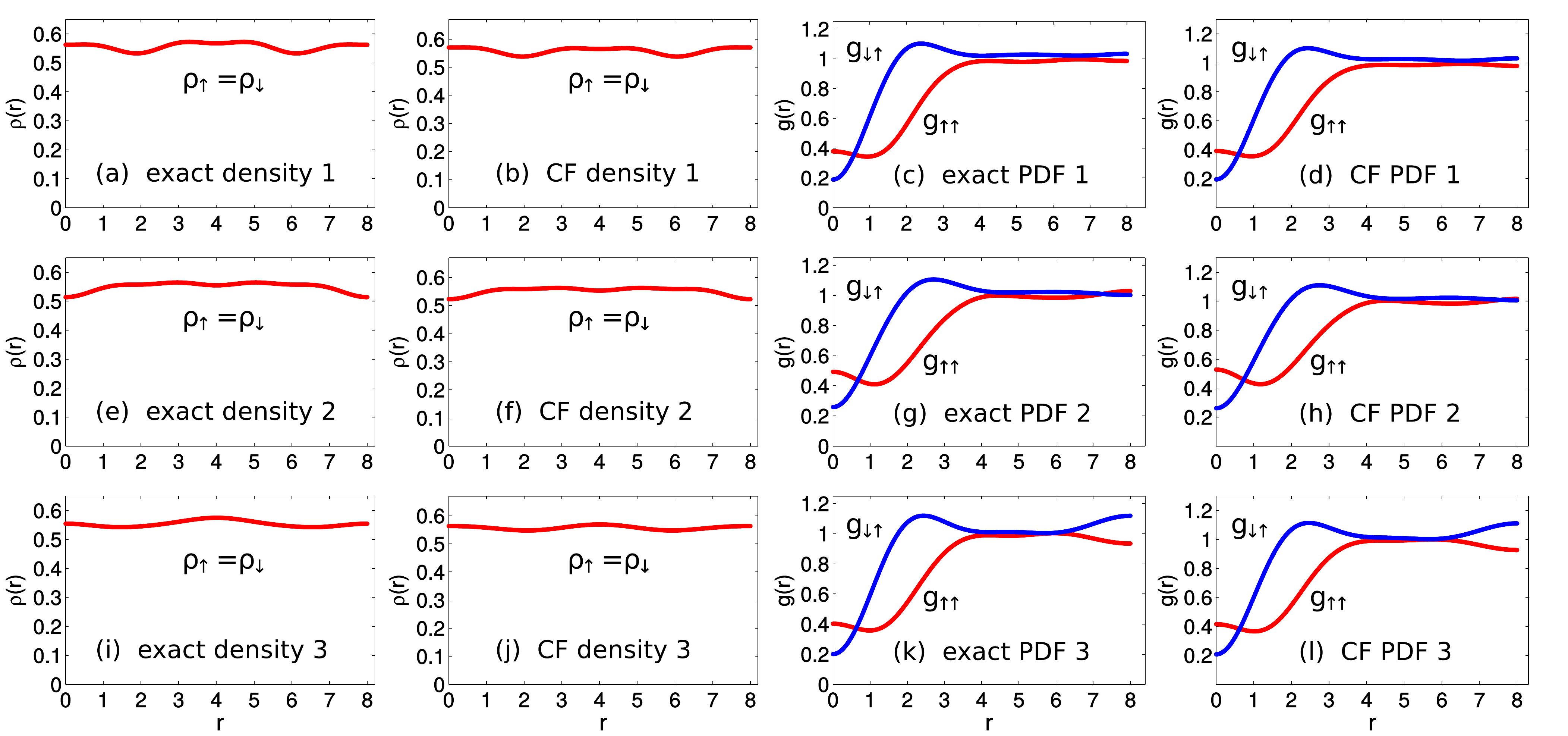}
\caption{(color online) Density profiles and pair distribution functions of the three lowest-energy exact eigenstates and the corresponding CF trial states (from top to bottom, labeled 1, 2 and 3) for the the $N=14$ system on sphere. The distance $r$ is measured in units of the magnetic length $\ell_B=\sqrt{{\hbar}c/eB}$ and the density values are given in units of $\ell^{-2}_B$. In panels (c), (d), (g), (h), (k) and (l), the two lines show $g_{\uparrow\uparrow}$ and $g_{\downarrow\uparrow}$ as indicated by the symbols in their vicinities.}
\label{Figure5}
\end{figure*}

We now discuss some issues relating to the experimental feasibility of preparing and detecting the emergent Fermi sea. It is straightforward to create two-component bosons, by imposing a coherent rotation that flips half of the bosons into a different internal state (along with an additional decoherence mechanism, such as a magnetic field gradient~\cite{Gupta} or an optical field that produces spontaneous photon scattering), or by optical pumping followed by evaporative cooling. The condition of an SU(2) invariant interaction requires that the atoms have the same scattering length in all channels, which can be very well achieved using Rb or Na. One limitation of the rotating BEC experiment~\cite{Gemelke} is that only a small number of particles can enter the FQH regime and the system cannot get very close to the ``centrifugal limit" $\Omega=\omega_c$. The filling factor $\nu=1$ is comfortably far from the centrifugal limit which occurs at $\nu=2/3$ for a two component system.

The cold atom experiments allow one to directly access some information of the wave function. In particular, one can measure the density profile by time of flight expansion and the pair distribution function (PDF) by either photo-association creation of molecules (which gives the short range behavior) or counting statistics (which in principle can give any equal time correlation function).  For a given state $|\Phi\rangle$, the PDF $g_{\sigma\tau}({\mathbf r}_1,{\mathbf r}_2)$ is defined as
\begin{eqnarray}
g_{\sigma\tau}({\mathbf r}_1,{\mathbf r}_2)=\frac{\langle\Phi| \psi^\dagger_{\sigma}({\mathbf r}_1) \psi^\dagger_{\tau}({\mathbf r}_2) \psi_{\tau}({\mathbf r}_2) \psi_{\sigma}({\mathbf r}_1) |\Phi\rangle}{\langle\Phi| \psi^\dagger_{\sigma}({\mathbf r}_1) \psi_{\sigma}({\mathbf r}_1) |\Phi\rangle \langle\Phi| \psi^\dagger_{\tau}({\mathbf r}_2) \psi_{\tau}({\mathbf r}_2) |\Phi\rangle}
\nonumber
\end{eqnarray}
where $\sigma,\tau$ are spin indices and $\psi^\dagger_{\sigma}({\mathbf r})$ [$\psi_{\sigma}({\mathbf r})$] is the creation (annihilation) operator for a boson at position ${\mathbf r}$. The density profiles $\rho$ and the PDFs $g_{\sigma\tau}({\mathbf r})$ of the exact ground eigenstate and the CF trial state for the $N=12$ system on disk are shown in Fig.~\ref{Figure4}. For the PDFs, we have defined ${\mathbf r}={\mathbf r}_1-{\mathbf r}_2$ and taken ${\mathbf r}_2$ to be the origin, so $g$ is only a function of $r=|{\mathbf r}|$ due to rotational invariance. The difference in the density profiles implies that the Fermi sea trial wave function is not very accurate for the edge physics. However, a comparison of the PDFs shows that the CF Fermi sea captures the bulk inter-particle correlations accurately. At the same time, the PDF also serves as a caveat against overextending the analogy between the CF Fermi sea of bosons and a spin-singlet Fermi sea of free fermions. The $g_{\uparrow\uparrow}$ curve shows that there is a large exchange correlation hole for bosons with the same spin (as expected), but it does not vanish at ${\mathbf r}=0$. Furthermore, bosons with different spins are also correlated, to be contrasted with the $g_{\uparrow\downarrow}({\mathbf r})$ of a Fermi sea of free fermions which does not depend on ${\mathbf r}$.

In addition to the density profiles and PDFs for the states on disk, we have also calculated these quantities for the states on sphere, which should better reflect the PDFs in the thermodynamic limit. One issue about the calculation on sphere is that most of the states presented in Fig.~\ref{Figure1} are not uniform states ({\em i.e.} they do not have $L=0$), and consequently the PDFs depend on the choice of both ${\mathbf r}_1$ and ${\mathbf r}_2$. For definiteness, we choose ${\mathbf r}_2$ to be at the north pole of the sphere and define $r$ as the chord distance between ${\mathbf r}_1$ and ${\mathbf r}_2$. The density profiles $\rho$ and the PDFs $g_{\sigma\tau}({\mathbf r})$ of the three lowest-energy exact eigenstates and the corresponding CF trial states for the $N=14$ system on sphere are shown in Fig.~\ref{Figure5}. We can see that variations in density are weak, the PDFs do not depend significantly on which low energy state is chosen, and have a similar behavior as that seen for the CF Fermi sea state in a rotating trap. 

\section{Conclusions}

The finite system studies presented above strongly support the existence of an SU(2) singlet Fermi sea at $\nu=1$, but are not definitive. To reveal the nature of the state at $\nu=1$, several decisive experimental manifestations of this state may be envisioned by analogy to those that led to the confirmation of a CF Fermi sea at half filling of electrons in the LLL~\cite{Willett,Kang,Goldman2,Kukushikin}. To begin with, the state will have no gap and no superfluidity no matter how small the temperature. It may be possible to measure the zero effective magnetic field by throwing an excitation off center and monitoring its oscillations in the trapping potential, which should not feel any Lorentz force in the rotating frame. Moving slightly away from $\nu=1$ should produce a Lorentz bending which is determined by the small effective magnetic field sensed by the particle as well as its Fermi velocity, a central property of the Fermi sea. When the filling factor is moved away from $\nu=1$, IQH states of composite fermions should occur and produce FQH states at $\nu=n/(n{\pm}1)$~\cite{Wu2}. 

The wave functions in Eq.~(\ref{TwoJainCF}) may also be interpreted in terms of ``partons"~\cite{Parton}. In this view, the boson is decomposed into two fictitious fermions (partons) and one species forms an IQH state while the other a Fermi sea. The patrons must of course be stitched together at the end to produce the physical bosons. This interpretation bears some similarity to the Bose metal phase discussed in Refs.~\cite{Mishmash,Jiang} in connection with the non-Fermi liquid ``strange metal" phase of cuprate superconductors. 

In conclusion, our calculations present strong evidence that two-component bosons in the LLL form an SU(2) singlet CF Fermi sea at $\nu=1$. Furthermore, advances in cold atom experiments provide unique opportunities for creating and studying this state, which represents extremely complex correlations between bosons but exhibit many properties that we associate with an ordinary Fermi sea. 

\section*{Acknowledgement} 

We are grateful to N. Gemelke and G. J. Sreejith for insightful discussions, and the authors of the DiagHam package for sharing their programs, especially N. Regnault. The work at Penn State was supported by the U.S. Department of Energy, Office of Science, Basic Energy Sciences, under Award No. DE-SC0005042 and the work at MPQ by the EU project SIQS.  We acknowledge the computing resources of the Research Computing and Cyberinfrastructure at Pennsylvania State University which is in part funded by the National Science Foundation grant OCI-0821527. 

\begin{appendices}

\section{CF Fermi Sea of One-Component Bosons}

The wave function describing the CF Fermi sea for a one-component system is
\begin{eqnarray}
\Psi^{\rm CF-FS}_{\rm 1-comp}(\{z\}) = {\cal P}_{\rm LLL} \left[ \Phi^{\rm FS}_{\rm 1-comp}(\{z\}) J(\{z\}) \right] 
\label{OneJainCF}
\end{eqnarray}
which can be understood in the same way as Eq.~(2) in the main text: the Jastrow factor $J=\prod_{i>j=1}^N(z_i-z_j)$ implements vortex attachment, $\Phi^{\rm FS}_{\rm 1-comp}(\{z\})$ is the wave function for spin-polarized Fermi sea in zero magnetic field, and ${\cal P}_{\rm LLL}$ projects the wave function to the LLL. 

\begin{figure}
\includegraphics[width=0.5\textwidth]{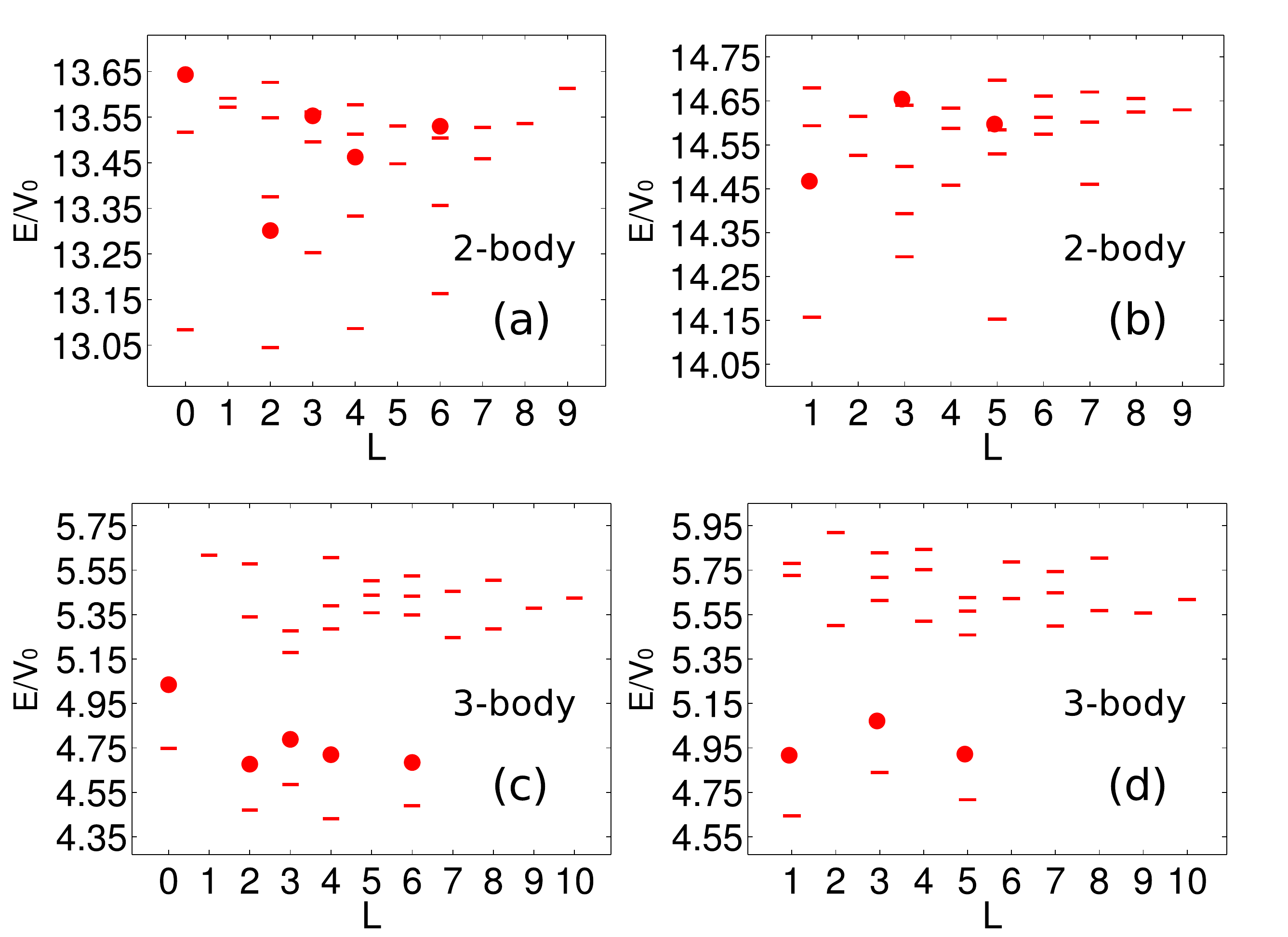}
\caption{Energy spectra of one-component bosons at $\nu=1$ on sphere (dashes). The system parameters are $(N,2Q)=(13, 12)$ in panels (a) and (c) and $(N, 2Q)=(14, 13)$ in panels (b) and (d). The panels (a) and (b) are for the two-body interaction $H^{(2)}_{\rm 1-comp}$ with $C^{(2)}_2=0.2$, while (c) and (d) are for the three-body interaction $H^{(2)}_{\rm 1-comp}$ with $C^{(3)}_2=0.2$. The dots show the energies of the wave functions $\Psi^{\rm CF}_{\rm 1-comp}$ with respect to the $H^{(2)}_{\rm 1-comp}$ in (a) an (c) and $H^{(2)}_{\rm 1-comp}$ in (b) and (d). The zeroth two-body or three-body pseudopotentials are chosen to be one and used as units of the energy values.}
\label{FigureA1}
\end{figure}

As mentioned above, for two-body contact interaction, composite fermions are produced but they do not form a Fermi sea but rather a paired Pfaffian state. One can ask if it is possible to obtain a CF Fermi sea state by varying the interaction, by making it longer ranged. We parametrize the interaction in the pseudopotential representation using the following Hamiltonian
\begin{equation}
H^{(2)}_{\rm 1-comp} = \sum_{ij}  \left[ P_{ij}(0) + C^{(2)}_2 P_{ij}(2) \right] 
\end{equation}
where $P_{ij}(L)$ projects out a pair of particles $i,j$ with relative angular momentum $L$, and $C^{(2)}_L$ is the interaction energy of two particles in the state with relative angular momentum $L$. The term $P_{ij}(0)$ is due the contact interaction and the term $P_{ij}(2)$ can be generated using longer-range interaction ({\em e.g.} dipole-dipole interaction between particles with permanent dipole moments). For a one-component system on sphere, the many-body eigenstates can be labeled by their total orbital angular momentum values. From exact diagonalization results at many different values of $C^{(2)}_2$, we find that the best match between the exact eigenstates and the CF Fermi sea trial wave function in Eq.~\ref{OneJainCF} occurs at $C^{(2)}_2=0.2$ as shown in Fig.~\ref{FigureA1} [panels (a) and (b)] and Table~\ref{TableS1}. The low energy band of states is consistent with that expected from the CF Fermi sea physics, but the quantitative comparison is far from convincing. 

Chung and Jolicoeur~\cite{Chung} have considered bosons interacting with a dipolar interaction. They have shown that at $\nu=1/3$ the bosonic system is best described by a Fermi sea of composite fermions made from the binding of bosons and three vortices.

Another possible realization of a Fermi sea is suggested from the following observation for fermions, the Pfaffian state at $\nu=1/2$ is the exact zero energy state for an appropriate three body interaction. It was found in Ref. \onlinecite{Wojs} that when the longer range part of the three body interaction is turned on, the Pfaffian state yields to a CF Fermi sea. The bosonic Pfaffian is exact for the three body contact interaction and one may  ask what happens when the three-body interaction becomes longer ranged. To address this, we write the Hamiltonian in the pseudopotential representation as
\begin{equation}
H^{(3)}_{\rm 1-comp} = \sum_{ijk}  \left[ P_{ijk}(0) + C^{(3)}_2 P_{ijk}(2) \right] 
\end{equation}
where $P_{ijk}(L)$ projects out a triple of particles $i,j,k$ with relative angular momentum $L$ and $C^{(3)}_L$ is its energy. (For small enough $L$, there is only one triplet state.) We find that longer-range three-body interaction indeed suppresses the pairing of composite fermions and leads to a CF Fermi sea. A comparison between the exact eigenstates of $H^{(3)}_{\rm 1-comp}$ with $C^{(3)}_2=0.2$ and the CF Fermi sea trial wave functions is shown in Fig.~\ref{FigureA1} [panels (c) and (d)] and Table~\ref{TableS1}. The reasonably good agreement between the energies and the high overlaps suggest that a suitable three-body interaction can produce a CF Fermi sea. We have also found that the qualitative and quantitative match persists over a wide range of $C^{(3)}_2$, indicating that no substantial fine tuning of parameters is required. While proposals have been made to engineer three-body interactions between bosons in cold atom systems~\cite{Daley1,Mazza,Mahmud,Daley2}, an experimental realization of this interaction is likely to be significantly more challenging than the two body contact interaction considered in the main text. 

\begin{table}
\centering
\begin{tabular}{|c|ccccccc|}
\hline
        &       &       &       &  $L$  &       &       &        \\\hline
Figure  &   0   &   1   &   2   &   3   &   4   &   5   &   6    \\\hline
 S1(a)  & 0.026 & -     & 0.846 & 0.615 & 0.608 & -     & 0.561  \\
        & (181)   &       & (779)   & (1024)  & (1359)  &       & 1900   \\\hline
 S1(b)  & -     & 0.750 & -     & 0.142 & -     & 0.570 & -      \\
        &       & (1300)  &       & (2944)  &       & (5174)  &        \\\hline 
 S1(c)  & 0.917 & -     & 0.943 & 0.954 & 0.922 & -     & 0.962  \\
        & (181)   &       & (779)   & (1024)  & (1359)  &       & 1900   \\\hline
 S1(d)  & -     & 0.928 & -     & 0.950 & -     & 0.960 & -      \\
        &       & (1300)  &       & (2944)  &       & (5174)  &        \\\hline  
\end{tabular}
\caption{The overlaps of the CF trial states with the corresponding exact eigenstates shown in Fig.~\ref{FigureA1}. $L$ is the orbital angular momentum and ``$-$" means that there is no trial state in that sector. The total number of linearly independent $L$ multiplets is given in parentheses below each overlap.}
\label{TableS1}
\end{table}

\end{appendices}

\end{document}